\documentclass[%
reprint,
twocolumn,
 amsmath,amssymb,
 aps,
pra,
 showpacs
]{revtex4-1}
\usepackage{amsmath}
\usepackage{mathrsfs}
\usepackage{graphicx}
\usepackage{dcolumn}
\usepackage{bm}
\usepackage{ulem}
\usepackage{gensymb}
\usepackage[
margin=.75in,
]{geometry}

\begin{document}
\title{Quantum mechanical study of high-order harmonic generation from HeH$^{2+}$ molecule in homogeneous and plasmonic-enhanced laser Fields}

\author{Nehzat Safaei}

\email {Corresponding author: safaei@ut.ac.ir}

\affiliation{
Department of Physical Chemistry, School of Chemistry, College of Science, University of Tehran, Tehran, I. R. Iran
 }

\begin{abstract}

High-order harmonic generation (HHG) from molecular ion HeH$^{2+}$ in two initial states of $1s\sigma$ (ground state) and $2p\sigma$ (first excited state) is investigated theoretically, in homogeneous and plasmonic-enhanced laser fields. The electron ionization and the enhancement of HHG yield in ground and first excited states, as the initial states, is studied and a strong orientation effect for ionization of electron in $2p\sigma$ state is discovered which it can be used to reach a longer lifetime for the electron in the first excited state during the high harmonic emission. In order to investigate effect of plasmonic field on the cutoff position of HHG spectra, first the enhancement of laser field in a nano-antennae is calculated and then the interaction of the enhanced field with HeH$^{2+}$ molecule is studied by solving the time-dependent Schr\"{o}dinger equation. According to the results, when HeH$^{2+}$ molecule, in the initial state of $2p\sigma$, is irradiated by plasmonic polarized laser field, with the polarization normal to the molecular axis, the HHG yield and cutoff position enhance while the excited state has a long lifetime during the HHG process. 
\end{abstract}
\maketitle
\section{Introduction}
High-harmonic generation (HHG) is an extremely nonlinear nonperturbative response of atoms and molecules to strong laser fields, which provides us an important tool to investigate ultrafast electronic dynamics [1-2]. When atoms and molecules are subject to intense laser radiation, new phenomena appear such as HHG and above threshold ionization (ATI) [3-5]. In particular, HHG has become a very interesting topic, since it is the most reliable way to get into the coherent light sources from spectral range of ultraviolet to extreme ultraviolet.
HHG mechanism can be interpreted by a semiclassical three-step model [6-8]. In this model, when atoms and molecules are exposed to the intense laser fields, the outer shell electron  tunnels through the Coulomb barrier, as a consequence of the nonperturbative interaction with the coherent electromagnetic radiation, and then the released electron is accelerated by the laser field and finally it may return to the parent ion due to a phase change of the electric field, followed by attosecond burst of electromagnetic waves emission.
Recent studies discovered some novel effects in HHG from asymmetric charged molecules such as HeH$^{2+}$ and LiH$^{3+}$. For example, Bian and Bandrauk [9] reported the orientation dependence of nonadiabatic effects in HHG from HeH$^{2+}$ ion, in ground state as the initial state,in short, intense laser pulses. Researches show that, in these asymmetric molecules with permanent dipoles, the excited states are localized states with a long lifetime. Hence the role of the excited states must be considered in molecular high-order harmonic generation (MHOHG) from asymmetric molecules, and the three-step model, which mainly takes the ground and continuum states into account, can not explain some properties of HHG from these molecules, such as the maximum cutoff energy [10]. Bian and Bandrauk [11] proposed a four-step model for molecular high-order harmonic generation from HeH$^{2+}$ molecule. By investigation of LiH$^{3+}$, BeH$^{4+}$ and HeH$^{2+}$, Feng and Chu [12] have shown that this excited state effect is a general character of the asymmetric molecules. On the other hand, previous investigations of HHG from atoms proved that excited states can enhance harmonic yields [13-14]. In this article, we use the simplest asymmetric molecular ion HeH$^{2+}$, which has a first excited state $2p\sigma$ with a comparably long lifetime, to investigate the orientation dependence of electron ionization and high harmonic emission from the molecule in initial sates of ground   and first excited states, to find an efficient way for harmonic generation.  Also, the plasmonic field, as one of the recent and important way to extend the cutoff energy of HHG, is implemented. The plasmonic field enhancement in nano-antennae has gained enormous interest and there has been a remarkable theoretical [15-19] and experimental activity [20-24] on this subject. The collective motion of confined electrons in a nanostructure exposed to an external electromagnetic field results in an enhanced field which can be calculated by solving Maxwell's equations. Recent experiments show that using a combination of engineered metal nanostructures and rare gases, HHG can be produced without using extra cavities or laser pumping to amplify the power of the input pulse [25], as the local electric fields enhances by more than 20 dB [26-27].
The numerical and semiclassical approaches to simulate the strong laser-matter interaction in particular high-order harmonic generation, is largely based on the dipole approximation for laser-atom/molecule interactions [15,16]. Within this assumption, the laser electric field (\textbf{E}(\textbf{r},t)) and its vector potential associated (\textbf{A}(\textbf{r},t)) are spatially homogeneous in the region where the electron dynamic takes place and spatial dependence of them is ignored i.e.\textbf{E}(\textbf{r},t) = \textbf{E}(t) and \textbf{A}(\textbf{r},t) = \textbf{A}(t). On the contrary, the field generated using plasmonic nanostructure is spatially dependent on a nanometer scale and cannot be described by this assumption.\\
\hspace*{2 mm}In present work, cutoff extension of HHG from HeH$^{2+}$ molecule in plasmonic-enhanced fields is calculated for two initial states of $1s\sigma$ and $2p\sigma$ and compared to the results from homogeneous fields.
The rest of the paper is organized as follows. In the next section (Sec. II), structure of the nano-antennae and calculation of the field enhancement are presented. In section  III, theoretical method  which is based on the time-dependent Schr\"{o}dinger equation, is described. Results and discussions are presented in Sec. IV. The last part of the paper, Sec. V, provides a short summary and outlook.
\vspace*{-7 mm}
\section{Structure of nano-antennae}
The nano-antennae is formed by two identical triangular gold pad separated by an air gap g. As it is presented in Fig. 1(a) the structure of nano-antennae is characterized by three geometrical parameters. The curvature radii of the tips are set to be 4 nm in order to avoid nonphysical fields enhancement due to tip-effect. The spatial profile of the field-enhancement around the bow-tie is determined by the means of finite-difference time domain (FDTD) calculations (COMSOL Multiphysics)[15], which is shown in Fig. 1(b). According to the Fig. 1(b) the laser electric field peak amplitude is enhanced  by a factor more than 2.4 near the metal tips compared with the center value.\\
\begin{figure}[ht]
\begin{tabular}{l}
\centering
\resizebox{80mm}{80mm}{\includegraphics[bb= 0 0 470 349]{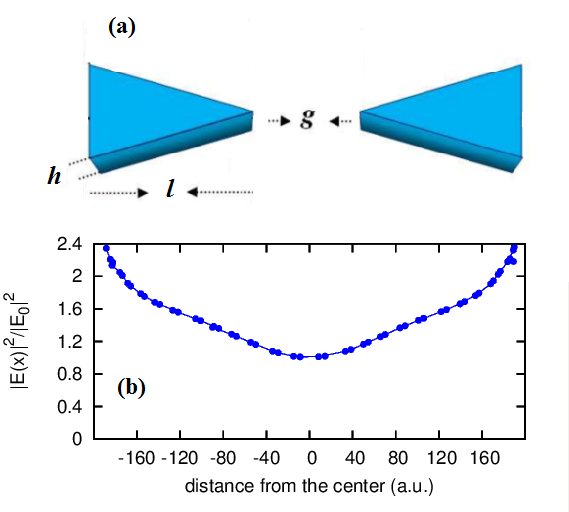}}
\end{tabular}
\caption{
\label{HHG} 
(Color online) Field enhancement in nanostructure. (a) Parameters of nano-antennae considered in simulations ( \textit{l}=100 nm, \textit{h}=40 nm, \textit{g}=20 nm). (b) The profile of the field enhancement in the gap, obtained from finite-difference time domain simulation. Circles are the actual determined enhancement and the solid line is the polynomial fitting.	}
\end{figure}
\section{Computational methods}
The interaction of HeH$^{2+}$ molecule with a linearly polarized laser field is described by the corresponding two-dimensional time-dependent Schr\"{o}dinger equation (TDSE) [10-11] as (atomic units are used throughout the paper.) 
\begin{eqnarray}\label{eq:11}
  i \frac{\partial \psi(x,y,t)}{\partial t}=[{\widehat {H}_0(x,y)}+\widehat H(x,y,t)]\,\psi(x,y,t),
\end{eqnarray}
where the unperturbed Hamiltonian $H_0$ of the system is given by                               
\begin{eqnarray}\label{eq:12}
\widehat{H}_0(x,y)=-\frac{1}{2}\nabla_{x,y}^2+ \widehat{V}(x,y).
\end{eqnarray}
In the above equation, $\widehat{V}$ is the soft coulomb potential of HeH$^{2+}$ molecule with softening parameter $\beta$ as 
\begin{eqnarray}\label{eq:3}
 \widehat{V}(x,y)= \frac{-2}{\sqrt{\beta+(x-R/2)^2}}+\frac{-1}{\sqrt{\beta+(x+R/2)^2}},
\end{eqnarray}
to produce the real energy curve of ${1s\sigma}$ and match the ionization potential of HeH$^{2+}$ ion, i.e $E_{1s\sigma}$ = -2.25 a.u..
The interaction term in length gauge, in case of $\theta=0^{\circ}$ ($\theta=90^{\circ}$), is $\widehat H(x,y,t)= E(t,x).x$ ( $\widehat H(x,y,t)=(E(t,y).y$). The laser polarization is along the molecular axis in case of $\theta=0^{\circ}$ (parallel) and normal to the molecular axis for $\theta=90^{\circ}$ (perpendicular).
During the simulations an absorbing potential included to avoid unphysical reflections of the electron wavepacket at the boundaries.
The TDSE is solved using unitary split-operator method where an eleven-point finite difference method is used for calculating the first and second derivatives. The Crank-Nicolson method which expresses the exponential operator to the third order is used to handle the time propagation.
Simulation boxes are set to 200 a.u. $\times$ 200 a.u. and the adaptive grid spacing is set to 0.2 a.u. (near the center of the simulation box) and 0.5 a.u. (near the borders of the simulation box) in both directions. The corresponding time step is set to be 0.01 a.u.. For HeH$^{2+}$ molecule the $1s\sigma$ ground state is repulsive whereas the first excited state $2p\sigma$ is a bound state and has a minimum at $R$= 3.89 a.u.. In present paper the internuclear distance is fixed at $R$ = 4 a.u., near the first excited-state minimum.
\begin{figure}[ht]
\begin{tabular}{l}
\centering
\resizebox{80mm}{67mm}{\includegraphics[bb= 5 4 463 279]{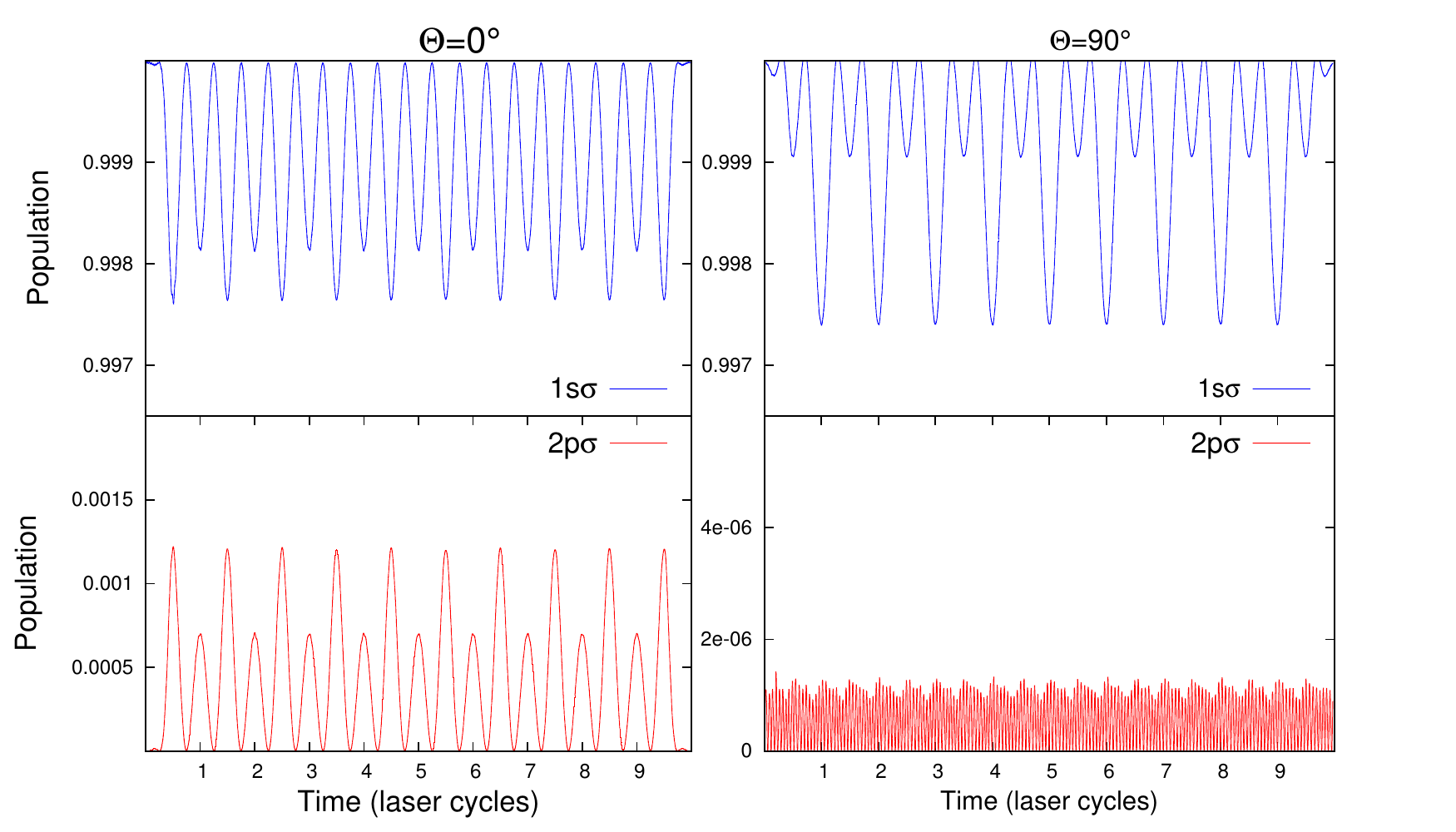}}
\end{tabular}
\caption{
\label{HHG} 
(Color online) Populations of ground ($1s\sigma$) and first excited ($2p\sigma$) states of HeH$^{2+}$ under a 10-cycle laser pulse of  800 nm wavelength ($\omega=0.057$ a.u.) and $I$=3$\times 10^{14}$ W/cm$^2$ intensity, with ground state as the initial wavepacket, for two relative orientations of the laser polarization and molecular axis, $\theta=0^{\circ}$ and $\theta=90^{\circ}$.}
\end{figure}
The excited state of the HeH$^{2+}$ molecule is determined numerically by Gramm-Schmitt orthogonalization method.
Based on the Ehrenfest theorem, the time-dependent dipole acceleration in case of $\theta=0^{\circ}$ can be read as [28]:
  \begin{eqnarray}\label{eq:3}
 d_A(t)=\langle \psi(x,y,t)\vert\, \widehat{x}.[\nabla \widehat{V}(x,y)+E(t,x)]\,\vert\psi(x,y,t)\rangle.
\end{eqnarray}
The HHG spectra are calculated as square of the windowed Fourier transform of dipole acceleration $d_A(t)$ in the direction of polarization of electric field as
 \begin{eqnarray}\label{eq:5}
  S(\omega)=\vert \frac{1}{\sqrt{2\pi}} \int_0^T d_A(t)\,H(t)\,exp[-i\omega t]\,dt\; \vert ^2,
\end{eqnarray}
where
\begin{eqnarray}\label{eq:6}
  H(t)= \frac{1}{2}[1-cos(2\pi \frac{t}{T})]
\end{eqnarray}
is the Hanning filter and $T$ is the total pulse duration. The time dependence of harmonics is obtained by Morlet wavelet transform of dipole acceleration $d_A(t)$ via:
\begin{eqnarray}\label{eq:7}
  &w(\omega,t)= \sqrt{ \frac{\omega}{\pi^\frac{1}{2}\sigma}}\times
 \nonumber \\
 &\int_{-\infty}^{+\infty}d_A(t)(t^\prime)exp[-i\omega (t^\prime-t)]exp[-\frac{\omega^2 (t^\prime-t)^2}{2\sigma^2}]dt^\prime.
\end{eqnarray}
\section{Results and Discussion}
In order to investigate the electron ionization and HHG spectra of HeH$^{2+}$ molecule in homogeneous and plasmonic-enhanced fields, Schr\"{o}dinger equation is solved numerically for two $\theta=0^{\circ}$ and $\theta=90^{\circ}$ orientations. The peak intensity of laser pulse is considered $I$=3$\times 10^{14}$ W$/$cm$^2$ and the wavelength is $\lambda=800$ nm ($\omega=0.057$ a.u.). The laser pulse has a trapezoidal shape, $f(t)$, with 0.5 cycles ramp on, 9 cycles constant, and 0.5 cycles ramp off.  
Interaction of the laser field with electron, in the case of $\theta=0^{\circ}$, can be written as:
 \begin{eqnarray}\label{eq:7}
 E(x,t) = E_0\, f(t)\,cos(\omega t)\,( 1 + g(x)).
\end{eqnarray}
 In the above equation, $g(x)$ is $\sum{ b_j x^j} $ and $b_j$ coefficients determined by fitting of data obtained from FDTD simulations, which they are set to be zero in case of homogeneous field.\\
\begin{figure}[ht]
\begin{tabular}{r}
\centering
\resizebox{80mm}{70mm}{\includegraphics[bb=9 6 200 203]{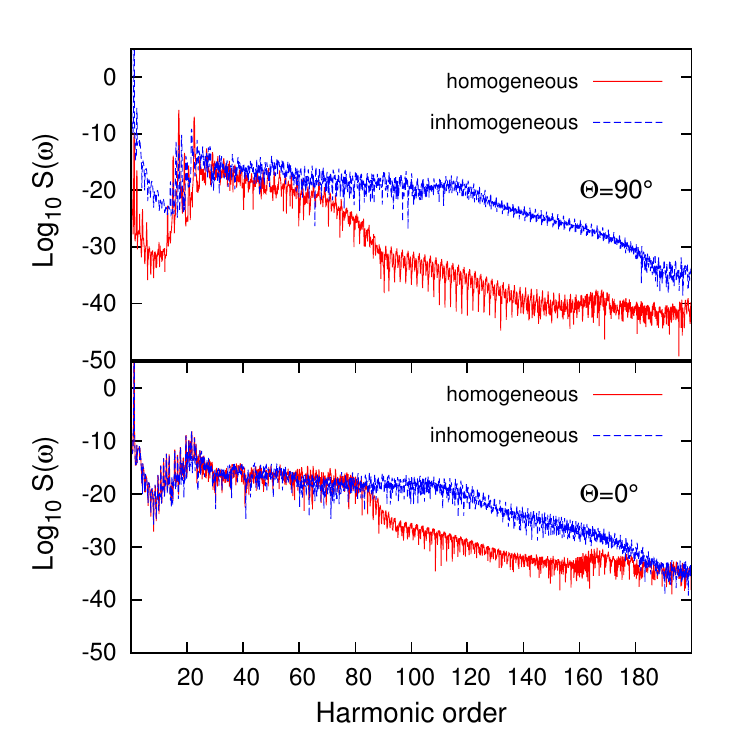}}
\end{tabular}
\caption{
\label{HHG} 
(Color online) High-order harmonic spectra produced by HeH$^{2+}$ under a 10-cycle laser pulse of 800 nm wavelength ($\omega=0.057$ a.u.) and $I$=3$\times 10^{14}$ W/cm$^2$ intensity, with ground state as the initial wavepacket, for two relative orientations of the laser polarization and molecular axis, $\theta=0^{\circ}$ and $\theta=90^{\circ}$.}
\end{figure}
\hspace*{0.4 mm}First, the interaction of HeH$^{2+}$ molecule, in initial state of $1s\sigma$, with homogeneous laser field is simulated. The calculated populations of $1s\sigma$ and $2p\sigma$ states, during the simulation, are presented in Fig. 2 as a function of laser cycle, for both $\theta=0^{\circ}$ and $\theta=90^{\circ}$ orientations. As it can be seen, when polarization of laser field is along to the molecular axis, the population changes of $1s\sigma$ and $2p\sigma$ states in every half cycle indicate the enhanced excitation (EE) and this shows that, $2p\sigma$ state could pay a key role in the HHG process. As it could be seen from Fig. 2, in contrast with parallel orientation, when polarization of laser field is perpendicular to the molecular axis, $\theta=90^{\circ}$, no significant change is observed in the population of $2p\sigma$ state, while the population of $1s\sigma$ state has the same pattern as $\theta=0^{\circ}$ orientation, which indicates the role of the next higher excitation states. 
\begin{figure*}[ht]
\begin{center}
\centering
\resizebox{140mm}{90mm}{\includegraphics[bb=9 4 487 212]{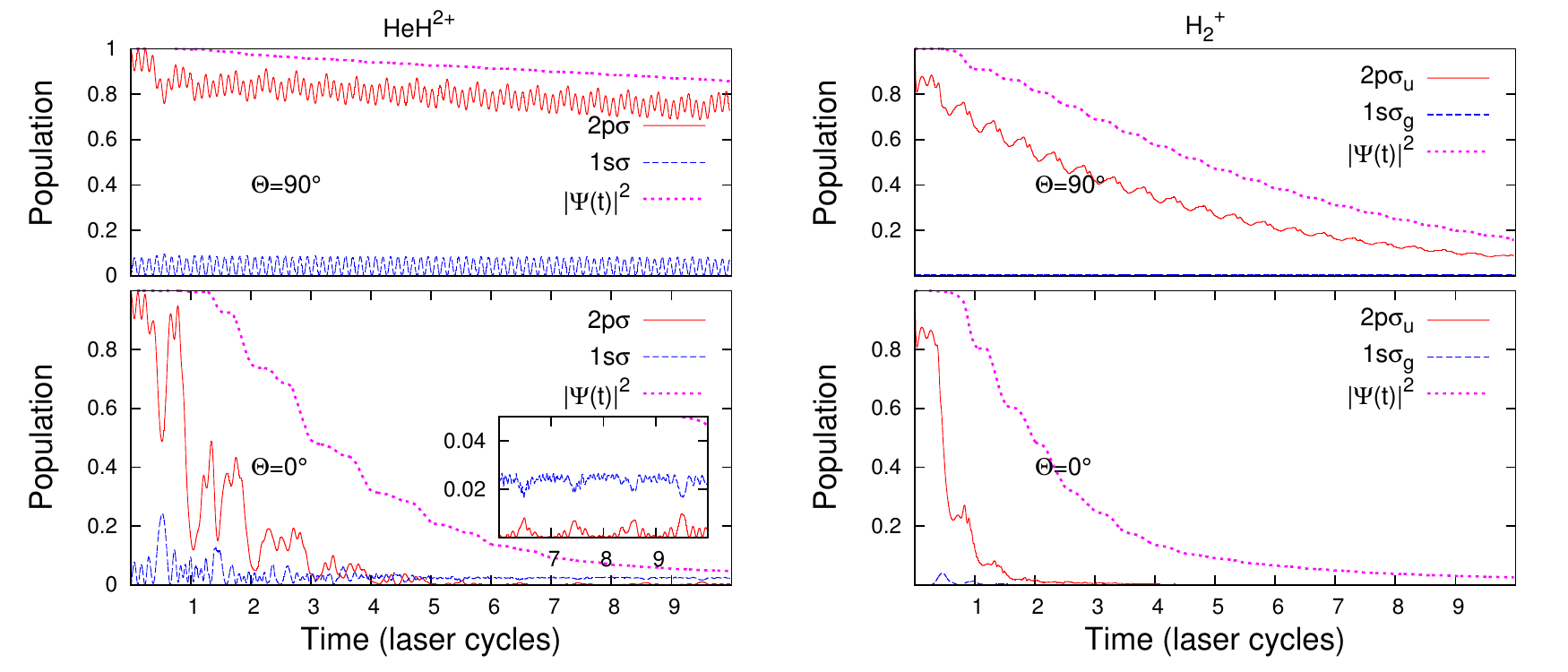}}
\caption{
\label{HHG} 
(Color online) Populations of ground and first excited states and electron probability density (norm) of HeH$^{2+}$ and H$2^{+}$ molecules as a function of time, under a 10-cycle laser pulse of 800 nm wavelength ($\omega=0.057$ a.u.) and $I$=3$\times 10^{14}$ W/cm$^2$ intensity, with first excited state as the initial wavefunction, for two relative orientations of the laser polarization and molecular axis, $\theta=0^{\circ}$ and $\theta=90^{\circ}$.}
\end{center}
\end{figure*}
As pointed out by Kamta et al. [29], for $F<0$ (the electric field amplitude $\vert F\vert$ corresponds to the peak laser intensity), the electric field shifts the energy of the ground and first excited states upward $+\vert F\vert R/2$ and downward $-\vert F\vert R/2$ respectively, which results in small energy gap between dressed states, and both EE and EI (enhanced ionization) occur. In contrary, for the $F>0$ part of the electric field, energy state of ground and first excited states shift downward and upward respectively, and the energy gap of these dressed states becomes large. Consequently, the electron excited and ionized at the first part ($F<0$) can transit to the ground state in the $F>0$ part and emit photons. As it is apparent from the Fig. 2, the periodic population change of ground state for $\theta=90^{\circ}$ orientation, is more than that for $\theta=0^{\circ}$, which one can conclude that, the energy gap of dressed ground and one of the higher excited states, in perpendicular orientation, is smaller than energy gap of dressed ground and first excited state in parallel orientation.
No difference was observed between population of states in homogeneous and inhomogeneous fields.\\
\hspace*{0.4 mm} The calculated HHG spectra for HeH$^{2+}$ molecule in initial state of $1s\sigma$, in both homogeneous and plasmonic fields, are presented in Fig. 3 for two $\theta=0^{\circ}$ and $\theta=90^{\circ}$ orientations.
From Fig. 3, one can see a resonance in HHG spectra when molecule was pumped by a laser pulse with polarization along the molecular axis, $\theta=0^{\circ}$, which is consistent with previous findings of Bian and Bandrauk [9]. HHG from an asymmetric molecule shows some differences compared to that from a symmetric system.
For asymmetric charged molecules with permanent dipoles, such as HeH$^{2+}$ molecular ion, lifetime of the excited states could be comparably long. The mean lifetime of the first excited state $2p\sigma$ of HeH$^{2+}$ is about 4 ns [30]. Linearly polarized intense laser field along the molecular axis pumps the system to the excited states and the resulting laser-induced electron transfer opens multichannel molecular high-order harmonic generation. In order to describe the observed resonance, Bian et al. [11] recommended a four-step model instead of the three-step semiclassical one. In this four-step model, first laser field pumps the electron from the ground state to the excited state and then part of the electron ionizes. The ionized electron accelerates in the laser field and finally recombines to the ground state and emits the radiation.
In contrast with $\theta=0^{\circ}$ orientation, there is no resonance for perpendicular orientation in Fig. 3, where the transition from the  $1s\sigma$ to $2p\sigma$ state is forbidden by selection rules.
As it can be seen in Fig. 3, there is an extension in cutoff position from 78 ($\theta=0^{\circ}$ orientation) and 70 ($\theta=90^{\circ}$ orientation)  in homogeneous field to 105 ($\theta=0^{\circ}$ orientation) and 118 ($\theta=90^{\circ}$ orientation) in inhomogeneous field.\\ 
\begin{figure}[t]
\begin{tabular}{l}
\resizebox{80mm}{80mm}{\includegraphics[bb= 9.198000 6.066000 199.399986 204.605994]{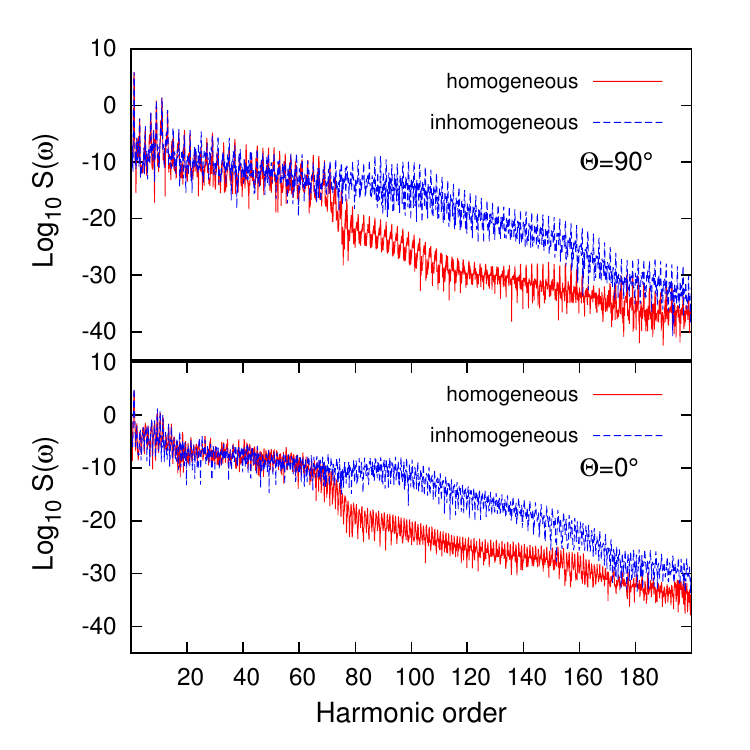}}
\end{tabular}
\caption{
\label{HHG} 
(Color online) High-order harmonic spectra produced by HeH$^{2+}$ under a 10-cycle laser pulse of 800 nm wavelength ($\omega=0.057$ a.u.) and $I$=3$\times 10^{14}$ W/cm$^2$ intensity, with first excited state as the initial wavepacket, for two relative orientations of the laser polarization and molecular axis, $\theta=0^{\circ}$ and $\theta=90^{\circ}$.}
\end{figure}
\begin{figure*}[ht]
\begin{center}
\centering
\resizebox{120mm}{100mm}{\includegraphics[bb=16.290000 4.104000 340.973990 359.999989]{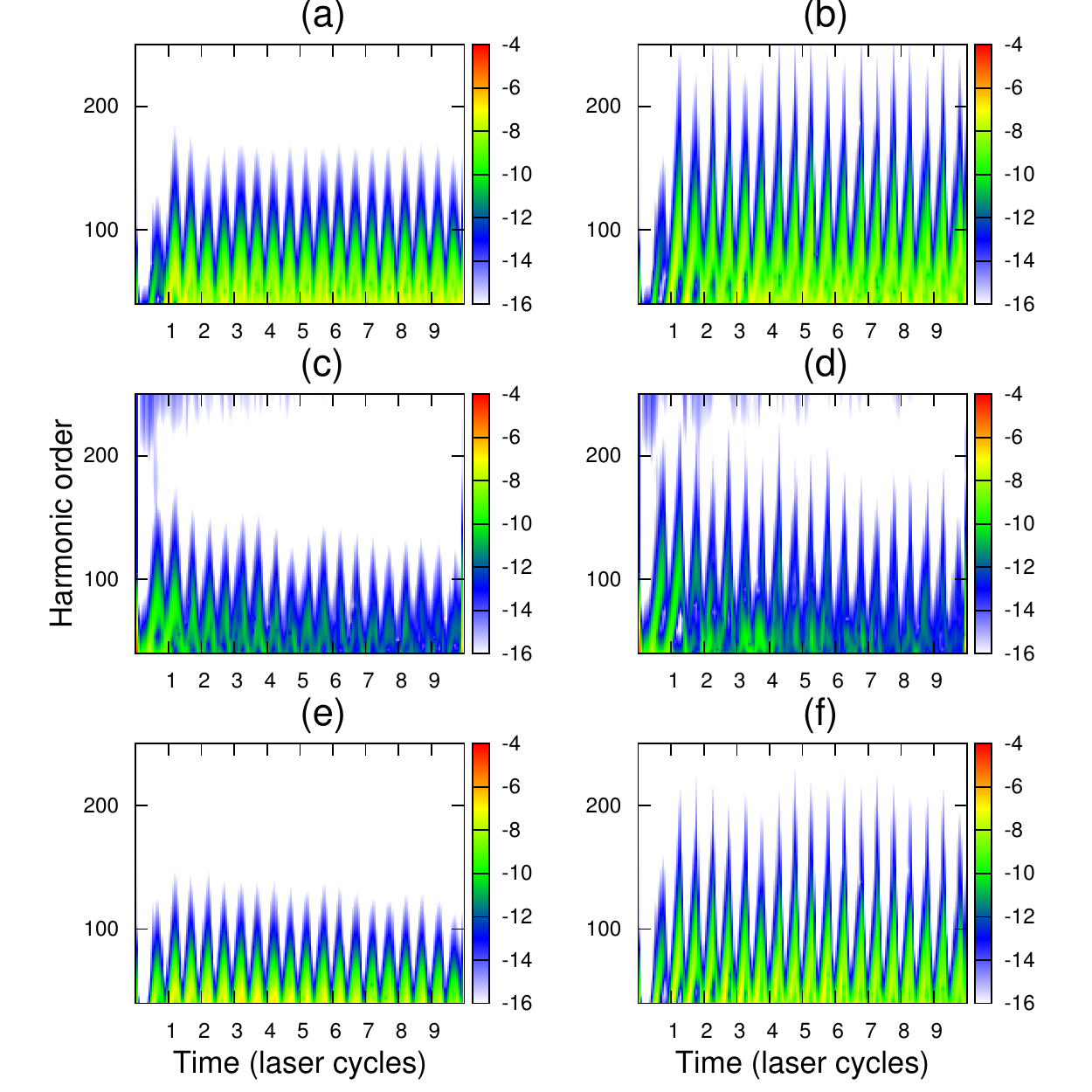}}
\caption{
\label{HHG} 
(Color online) The Morlet wavelet time profiles of dipole acceleration for (a) HeH$^{2+}$ molecule with initial state of $2p\sigma$ in homogeneous (b) and plasmonic-enhanced laser field, (c) HeH$^{2+}$ molecule with initial state of $1s\sigma$ in homogeneous (d) and plasmonic-enhanced laser field,  (e) H$_2^{+}$ molecule with initial state of $2p\sigma_u$ in homogeneous (f) and plasmonic-enhanced laser field. In all panels the relative orientation of the laser polarization and molecular axis is $\theta=0^{\circ}$, and the parameters of laser pulse are the same as in the previous figures. }
\end{center}
\end{figure*}
 \hspace*{0.4 mm}In this part, the electron ionization and the high-order harmonic emission form HeH$^{2+}$ molecule, in the initial state of $2p\sigma$, is investigated. The electronic probability density of $2p\sigma$ state is mainly localized on the H$^+$ core, while the  probability density of electron in $1s\sigma$ state concentrates on the He$^{2+}$ core.
The populations of $1s\sigma$ and $2p\sigma$ states, and the probability density of electron (norm), as a function of laser cycle, in homogeneous laser field, is illustrated in the left panel of Fig. 4 for both $\theta=0^{\circ}$ and $\theta=90^{\circ}$ orientations. For the sake of comparison, the norm and the population changes of ground state $1s\sigma_g$ and first excited state $2p\sigma_u$ of symmetric molecule H$_2^{+}$, in initial state of first excited state, in a laser field with same parameter as field applied for HeH$^{2+}$ molecule, is presented in right panel of Fig. 4.  
As it can be seen in Fig. 4, when  molecule is irradiated by linearly polarized laser field along the molecular axis, $\theta=0^{\circ}$, most of the electron, in both HeH$^{2+}$ and H$_2^{+}$, ionizes during the radiation. In case of asymmetric molecule HeH$^{2+}$, there is a population transformation between ground and first excited states in the last laser cycles, as it is shown in the closed window, which corresponds to the electron transfer between nucleus He$^{+2}$ and H$^+$. This population transformation between ground and first excited state in the initial state of $2p\sigma$, keeps the intensity of HHG higher than that for the initial state of $1s\sigma$. When polarization of laser field is normal to the molecular axis, $\theta=90^{\circ}$, in contrast with symmetric molecule H$_2^{+}$, the ionization rate of electron in asymmetric molecule HeH$^{2+}$ is much less than that for the case of $\theta=0^{\circ}$ orientation. This observed strong orientation effect for electron ionization could be used to reach a longer lifetime for electron in the first excited state, which is responsible for the enhancement in HHG yield, during the HHG process. Like the initial state $1s\sigma$, no difference was observed between populations of states, for initial state of $2p\sigma$, in homogeneous and plasmonic-enhanced field. Therefor, the extension of the cutoff in plasmonic fields, comes from the further acceleration of the ionized electron due to the gradient of the laser field. 
The HHG spectra of HeH$^{2+}$ in $2p\sigma$ initial state, in homogeneous and plasmonic-enhanced fields, is presented in Fig. 5 for $\theta=0^{\circ}$ and $\theta=90^{\circ}$ orientations. As it could be seen, HHG from HeH$^{2+}$ molecule in initial state $2p\sigma$ has a higher intensity compared to that in $1s\sigma$ initial state, in both orientations. An extension in cutoff position from 65 in homogeneous field to 95 in plasmonic field is observed for both orientations.
The calculated time profiles of HHG emission from HeH$^{2+}$ in $1s\sigma$ and $2p\sigma$ initial states and $\theta=90^{\circ}$ orientation, in homogeneous and plasmonic-enhanced laser fields, are shown in Fig. 6 (a)-(d). The cutoff extension is obvious for HHG in inhomogeneous field for both initial states. Consistent with Fig. 5, when the HeH$^{2+}$ molecule is in the initial state of $2p\sigma$, one can see an enhancement in the HHG yield, in comparison with initial state of $1s\sigma$. As it is shown in the Fig. 6(a) and 6(b), in case of the initial state of $2p\sigma$, the high harmonic emission from HeH$^{2+}$, takes place during entire time of laser pulse. For comparison, the time profile of high harmonic emission from H$_2^{+}$ in the first excited state and $\theta=90^{\circ}$ orientation, in homogeneous and plasmonic-enhanced laser fields, is presented in Figs. 6(e) and 6(f) respectively. As one can see from the comparison of Figs. 6(a) and 6(b) with 6(e) and 6(f), HHG from HeH$^{2+}$ molecule has higher cuttof than H$_2^{+}$ molecule, during radiation.\\
\section{Conclusions and outlook}
The generation of high-order harmonics by HeH$^{2+}$ molecule, in two initial states of $1s\sigma$ and $2p\sigma$, is investigated in homogeneous and plasmonic-enhanced laser fields. The orientation dependence of electron ionization and high-order harmonic emission in both initial states is presented and compared for two relative orientations of the laser polarization and molecular axis, $\theta=0^{\circ}$ and $\theta=90^{\circ}$.  It is found that, when the molecule is irradiated by polarized laser field normal to the molecular axis, $\theta=90^{\circ}$, the excited state, which is responsible for the enhancement in HHG yield, has a longer lifetime than the case for $\theta=0^{\circ}$. HHG from $2p\sigma$ initial state has a higher intensity and lower cutoff compared to initial state of $1s\sigma$ in both orientations.
The cutoff extension of HHG from HeH$^{2+}$ molecule in a plasmonic field, made in a bowtie-shaped gold nano-antennae, is calculated and it is about 30 harmonics for the first excited state $2p\sigma$.
\section{Acknowledgments}
The author sincerely thanks Dr H. Ahmadi for valuable discussions.

\end{document}